# EPCI: A New Tool for Predicting Absolute Permeability from CT images

H. Sun, H. Al-Marzouqi, and S. Vega

ABSTRACT

A new and fast Matlab algorithm for predicting absolute permeability is presented. The developed tool relies on measuring the connectivity of pores in a given three-dimensional (3D) micro-CT rock image. An index of pore connectivity is introduced. After a calibration step, the developed index is used to estimate permeability in a variety of rocks with challenging pore structures (e.g. complex carbonate formations). The developed algorithm was tested on sandstone and carbonate rock samples. It offers large computational and memory savings when compared with algorithms based on the Lattice Boltzmann Method (LBM). Permeability estimates were, in general, in good agreement with laboratory measurements and numerical simulation results. Source code for computing the developed index along with an associated GUI panel are available online at https://github.com/cupbkust/EPCI.git

INTRODUCTION

In recent years, Digital Rock Physics (DRP) techniques have progressed at a rapid speed. These methods rely on imaging technologies to acquire 3D high-resolution representations of rock samples. The pore space and mineral matrix of natural rocks are digitized and then numerically simulated to obtain estimates for various macroscopic rock properties. Image volumes acquired using high-resolution micro-scale X-ray computed tomography (CT) are used to capture and visualize the three-dimensional pore geometry structure of reservoir rocks. A highly accurate 3D digital rock sample is used to extract the information of mineral composition and microstructure of rocks, by utilizing image-processing methods (Al-Marzouqi, 2018, Andra et al., 2013). Numerical simulations can then be performed on the digital rock sample (resulting digital rock model) to quantify the available pore space, fluid transport properties (e.g. absolute permeability) and other rock properties such as elastic modulus, and the formation resistivity factor (Saenger et al., 2016).

Several methods have been developed to calculate permeability from CT images. The lattice Boltzmann method (LBM) is commonly used for inferring the permeability of rocks from fluid simulations. The LBM models the fluid system as a group of discrete particles with varied velocities. These particles collide and stream over a discrete lattice at a mesoscopic level (Krüger, 2017). Permeability estimates obtained using LBM were shown to be in good agreement with standard laboratory measurements (Sun et al., 2017).

Dong and Blunt (2009) used pore network modeling (PNM) to predict network permeability of digital rocks. This method extracts a network representing the connected pores in the CT scan. The individual network elements are uniform ducts with circular, triangular or square cross-sectional shapes. Analytical formulas are then used to predict permeability values from the extracted network. The developed method was validated using experimental permeability values obtained from several rock samples. Different algorithms for extracting improved pore network models were proposed in the literature (e.g. Gostick et al., 2017; Yi et al., 2017).

Recently, deep neural networks were utilized for permeability estimation from 2D thin sections (Araya-Polo et al., 2018). The range of permeabilities for samples used in this paper was limited to 50-1100mD. Deep learning models require a large number of training samples and the estimates obtained by these networks are not easily interpretable.

Other methods for permeability estimation include the use of finite difference methods (FDM) (Øren and Bakke, 2002), pore-morphological modeling (Adalsteinsson et al., 2007), effective medium theory (Jurgawczynski, 2007), and the use of Minkowski functions (Vogel et al., 2010)

An important limitation in available tools used for permeability estimation is the high computational requirements. Large scale simulations can take days and often require the use of computer clusters. In addition, constraints in memory requirements limit the size of data samples. In this paper, we introduce a new and fast algorithm to predict absolute permeability of rock using 3D CT scans. Permeability estimates are obtained using a sequence of efficient binary operations and are found to be, in general, in good agreement with laboratory measurements and other numerical estimates.

ALGORITHM OVERVIEW

In this section, we present an overview of the developed Effective Pore Connectivity Index (EPCI) and the algorithm used to compute it. Computed index values are converted to permeability estimates using linear calibration.

**Effective Pore Connectivity Index (EPCI)**

The developed index depends on tracking the connectivity of pores within a rock sample. The developed algorithm first assumes that pores in the first slice are all filled with fluids. Next, it finds the pores to be filled with fluids in the second slice. A pore voxel in the second sliced is labeled to be filled with fluid if it is connected to a pore in the first slice. A pore is labeled connected if one of it is nine neighbors in the previous slice is a pore pixel. Assuming that image slices are stacked vertically, the nine neighbors include the pixel that is right above the pixel of interest and its eight adjacent neighbors. Figure 1 and Figure 2 show examples that illustrate the connectivity between two vertically adjacent slices. These figures show two slices of an image of size 3 by 3. In Figure 1, only the center pixel in the first slice is a pore (digit equals 1). In the second slice, all pixels indicate the presence of pores. According to our connectivity definition all the pores in the second image are connected to the first slice and will allow fluid to pass through them. If the first image did not contain pores then no pores in the second slice will be labeled as connected.

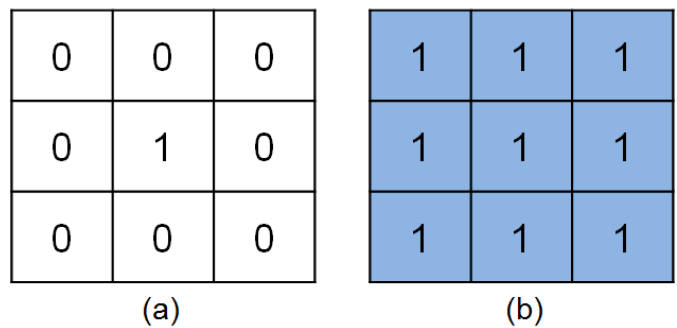

Figure 1 An example illustrating the operation of the developed pore connectivity index (a) first slice (b) second slice. Pores are labeled by the index 1. According to our connectivity definition all the pores in the second image are connected to the first slice.

In the second example, shown in Figure 3, only the upper left corner element of the first slice is labeled as a pore. The output of the algorithm will indicate the presence of 4 connected pores in the upper left corner of the image. The remaining pores are not connected.

|   |   |   |   |   |   |   |
|---|---|---|---|---|---|---|
| 1 | 0 | 0 |   | 1 | 1 | 1 |
| 0 | 0 | 0 |   | 1 | 1 | 1 |
| 0 | 0 | 0 |   | 1 | 1 | 1 |
| (a) |   |   |   | (b) |   |   |

Figure 2 A second example illustrating the operation of the developed pore connectivity index (a) first slice (b) second slice. Pores are labeled by the index 1. The four pixels in the upper left corner of the second slice are connected to the first slice.

The preceding connectivity computation is iterated to subsequent slices. Once we reach the last slice, the number of connected pores is used as an indicator of the permeability of the rock sample. A pore voxel in the last image slice is considered connected, when a feasible path for fluid flow exists from the first slice to the voxel. A rough estimate of permeability is obtained by counting the number of connected pores in the last slice. To reach a more reliable estimate the previous operation is repeated in the opposite direction (i.e. starting from the last slice).

Finally, we compute the EPCI value defined as the average number of connected pores found in the two computed directions.

**Algorithm description**

In what follows, we present a detailed description of the algorithm used in computing the connectivity index. The algorithm assumes that the input image is segmented. Denote each CT image slice as a binary function of pixel position $f(x, y)$. The j$^{th}$ image slice is given by $f_j(x, y)$. Assume that each image slice is of size $W \times H$ pixels and that the image volume consists of $Z$ slices.

Step 1 let j=1

Step 2 compute $f_{j,translate}$ as defined by the following equation:

$$f_{j,translate} \stackrel{\text{def}}{=} f_j(x,y) \mid f_j(x+1,y) \mid f_j(x+1,y+1) \mid f_j(x,y+1) \mid f_j(x-1,y+1) \mid$$
$$f_j(x-1,y) \mid f_j(x-1,y-1) \mid f_j(x,y-1) \mid f_j(x+1,y-1), \quad (1)$$

where '|' indicates the binary 'OR' operation. f$_{j,translate}$ shifts slice j to the eight different adjacent spatial positions and performs a logical OR operation between the original image and each of the eight resulting images.

Step 3 perform a logical "AND" operation between $f_{j,translate}$ and $f_{j+1}(x, y)$ and store the result in the matrix $C$. $C$ is a binary matrix of size $W \times H$ that tracks the connectivity of pores.

$$C \stackrel{\text{def}}{=} f_{j,translate} \& f_{j+1}(x,y), \tag{2}$$

where '&' is the logical "AND" operation.

Step 4 if $j$ is equal to the number of slices in the image volume $Z$, compute the connectivity index, as defined by

$$EPCI_{1 \to Z} = \frac{1}{W \times H} \sum_{w=1}^{W} \sum_{h=1}^{H} C, \tag{4}$$

and terminate the algorithm. Otherwise move to Step 5.

Step 5 compute $C_{translate}$ as defined by the following equation:

$$C_{translate} = C(x,y) \,|\, C(x+1,y) \,|\, C(x+1,y+1) \,|\, C(x,y+1) \; C(x-1,y+1) \,|$$
$$C(x-1,y) \,|\, C(x-1,y-1) \,|\, C(x,y-1) \,|\, C(x+1,y-1), \tag{3}$$

where '|' indicates the binary 'OR' operation. $C_{translate}$ shifts the connectivity matrix $C$ to the eight different adjacent spatial positions and performs a logical OR operation between the original image and each of the eight resulting images.

Step 6 increment $j$ by one and perform a logical "AND" operation between $C$ and $f_{j+1}(x,y)$ and update matrix $C$ with the new result, as follow:

$$C = C_{translate} \ \& \ f_{j+1}(x,y), \tag{5}$$

where '&' is the logical "AND" operation. Go back to step 4.

The preceding algorithm computes the EPCI value from Slice 1 to Slice Z. To get a more accurate estimate, we compute the EPCI value in the opposite direction and average the two results:

$$EPCI = (EPCI_{1 \to Z} + EPCI_{Z \to 1})/2, \tag{6}$$

Our experiments, presented in the next section, demonstrate a strong linear relationship between the EPCI index and rock permeability. Computing EPCI values for a sample of rocks with varying permeability values is required as a pre-processing step to get the parameters calibrating the relationship between the EPCI index and standard rock permeability values.

We would like to note that computation of pore connectivity between slices is performed using binary logical operations. The simplicity of the proposed tool and the use of binary logical operations make the developed permeability index more efficient than alternative algorithms frequently used in practice. Furthermore, the slice-by-slice incremental nature of the algorithm greatly reduces memory requirements. The algorithm is implemented using Matlab and the source code is available online.

## APPLICATIONS AND ANALYSIS

**Digital rock samples**

Nine sandstone samples and five carbonate samples were used to test the developed algorithm. Sandstone rock scans were downloaded from the website of Petroleum Engineering & Rock Mechanics group at Imperial College London (http://www.imperial.ac.uk/earth-science/research/research-groups/perm/research/pore-scale-modelling/micro-ct-images-and-networks/). Sandstone rock samples provided were provided by Saudi Aramco, Shell, Total, and Imperial College. They are of size 300×300×300 voxels and their image resolutions vary between 3.4 µm/voxel and 9.1 µm/voxel with an average resolution of 6.01 µm/voxel and standard deviation σ= 2.2. Laboratory estimates for permeability were not available for the Sandstone samples. We used estimates based on LBM and PNM simulations instead. LBM and PNM permeability estimates were obtained from Dong and Blunt (2009). In their paper, LBM used a D3Q19 model and bounce-back boundary condition to do the simulation and PNM used the maximal ball (MB) method to extract the pore network.

Carbonate samples were obtained from a field in the Middle East (https://figshare.com/s/d3816fda47ba7212cc48). Experimental porosity and permeability values, for the carbonate samples, were obtained using laboratory experiments. Porosity measurements were conducted by using a gas expansion helium porosimeter at ambient conditions to measure porosity by applying the gas transfer method (Boyle's Law), to determine the grain volume and calculate the pore volume. Permeability measurements were conducted using a steady state gas permeameter with an ambient confining pressure of 400 psi and Nitrogen as the flowing fluid. Darcy law was applied to calculate permeability (Mokhtar, 2014). Carbonate samples are of size 800×800×800 voxels and their image resolution is about 1.0 µm/voxel. Figure 3 shows the relationship between porosity and permeability for these samples. A good linear relationship for sandstone samples is observed, while for carbonate samples, there is no relationship between porosity and permeability, as the Pearson correlation coefficient value is only 0.05.

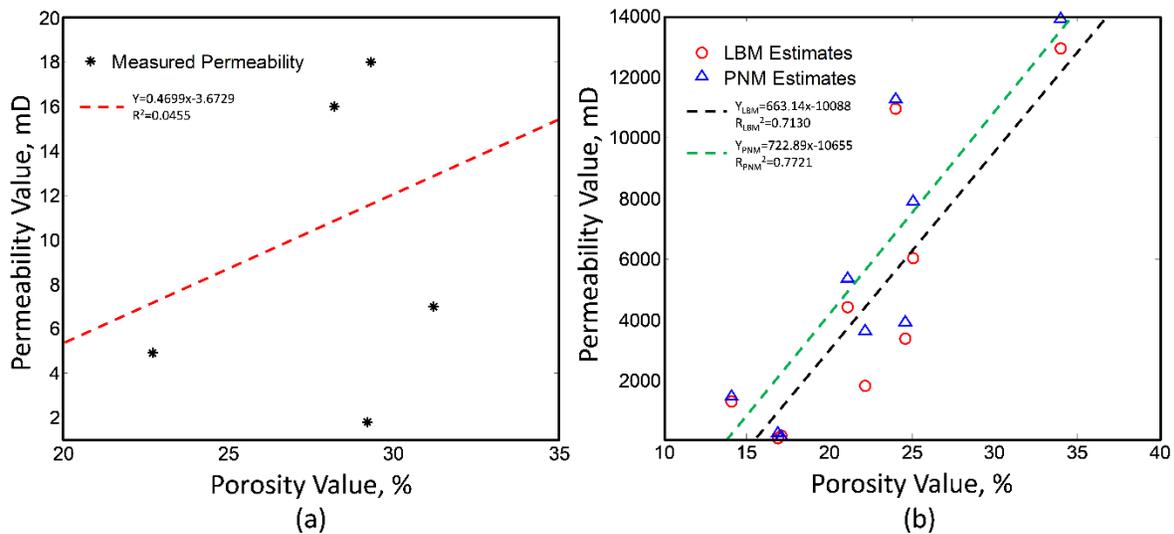

Figure 3 Relationship between porosity and permeability for carbonate samples (a) and sandstone samples (b).

**Noise removal and image segmentation**

Processing of images in DRP workflows typically involve a binary segmentation algorithm, where the acquired high-resolution images are converted to pores and grains. For carbonate rock samples, we performed non-local means (NLM) filtering on the acquired data samples to reduce noise levels. NLM filter operates on a non-local area by using a dissimilarity measure between a central region patch and neighboring patches in a searching window and it works well for removing Gaussian noise (Buades et al., 2005). Next, a marker-controlled watershed segmentation method is used to convert the smoothed image into a binary matrix where intensity values of 0 represent pores and values of 1 represent grains (Cristoforetti et al., 2008).

The downloaded sandstone samples are all segmented images. They were processed by a median filter and Otsu's method (Dong and Blunt, 2009). Two examples of segmented sandstone and carbonate rocks samples are shown in Figure 4.

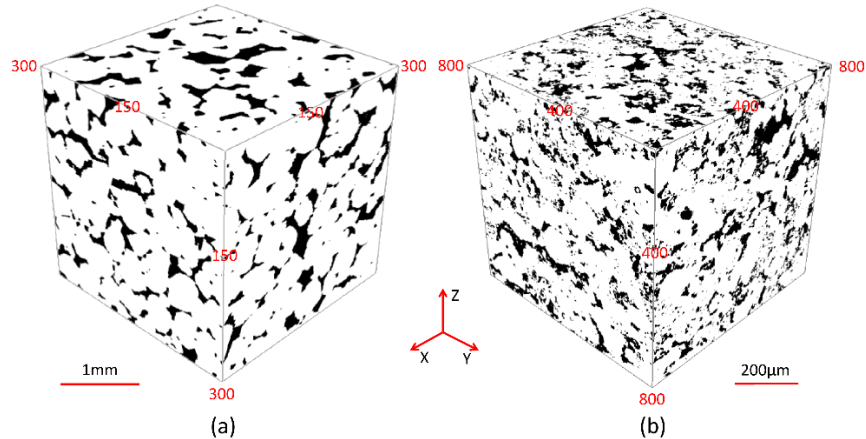

Figure 4 Examples for segmented 3D image of sandstone sample S1 (a) and carbonate sample C1 (b). White and black regions represent grains and pores, respectively

**Velocity field comparison**

We start by comparing velocity field data obtained used LBM simulations and the C matrix representing the connectivity information used to compute the EPCI index. Figure 5a shows a comparison between the LBM velocity fields in slice 150 in the first Sandstone sample (S1). In Figure 5b, EPCI's connectivity matrix C is shown for the same slice where black color indicates fluid flow pixels. The connectivity matrix shown is formed using the combination of forward and backward connectivity matrices $C_{150\ (1\rightarrow Z)}$ and $C_{150\ (Z\rightarrow 1)}$. As expected, the C matrix in EPCI highlights most fluid channels in LBM.

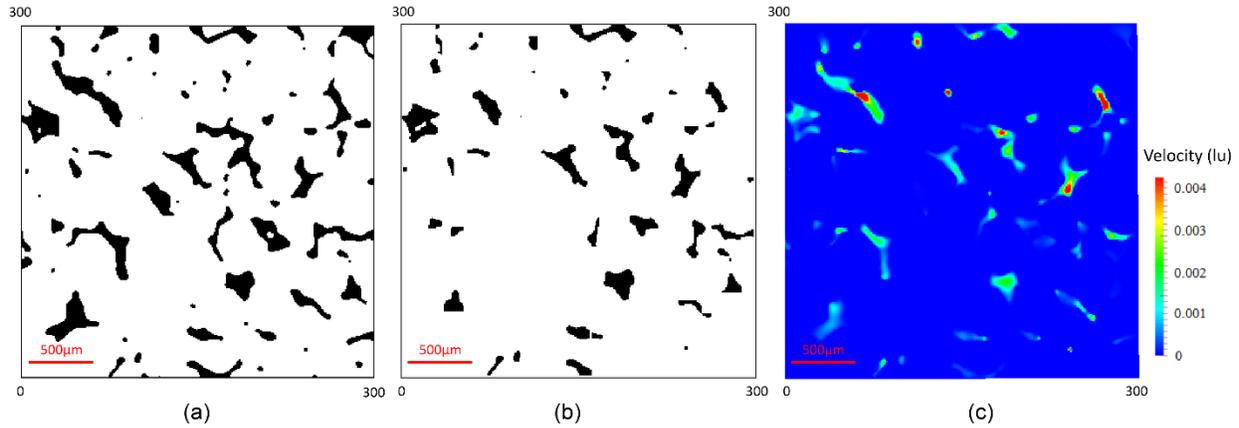

Figure 5 An example of slice comparison (Slice number 150 in S1) between LBM simulation and EPCI calculation: (a) original slice of S1; (b) EPCI slice containing the combination of connectivity matrices $C_{150}$ (1→Z) and $C_{150}$ (Z→1). Black indicates areas that the EPCI index considers connected to the first or last slice; (c) Velocity field (lattice unit, lu) obtained by LBM simulation.

**Estimating permeability**

The relationship between the developed index and permeability values was tested using the available digital rock samples. We used experimental lab values as a ground truth reference for carbonate rock samples. Permeability estimates of Sandstone sample images were computed using the Lattice Boltzmann method (LBM) and the Pore Network Model (PNM), as no experimental data was available.

The relationship between EPCI values and experimental data for carbonate samples and simulated permeability estimates obtained using LBM and PNM simulations for sandstone samples is illustrated in Figure 6. In Figure 6a, a strong linear relationship is observed between EPCI values and permeability measurements (see red dash line). The square of the Pearson linear correlation coefficient value $R^2$ is 0.92. In Figure 6b, a linear relationship between EPCI values and LBM/PNM predictions is also observed. $R^2$ values of 0.83/0.88 are observed between the EPCI and LBM estimates/network permeability estimates. Obtained correlation values significantly improve the porosity correlations shown earlier in Fig. 3.

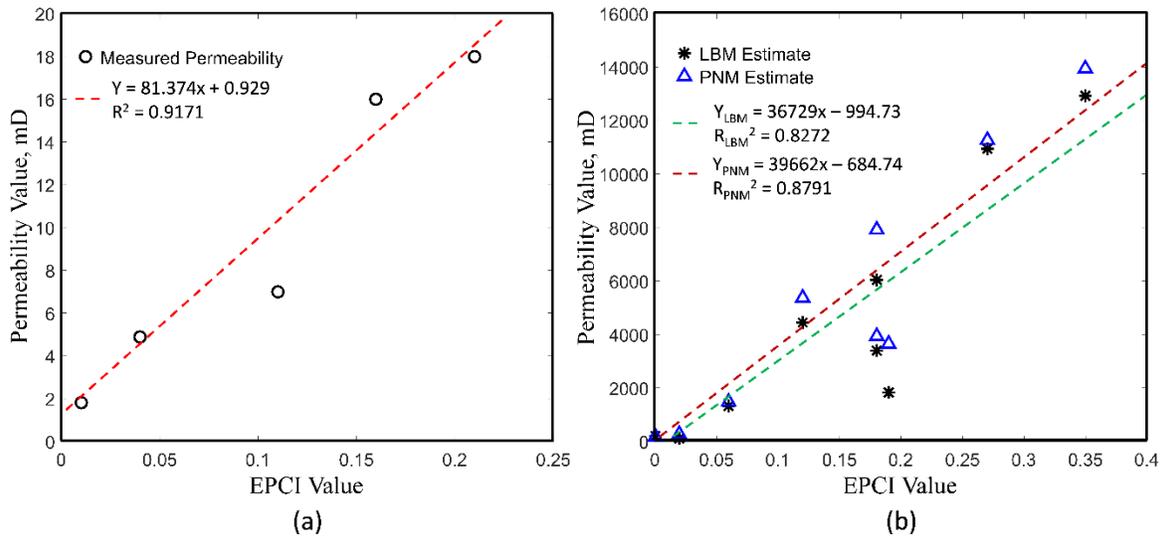

Figure 6 Plots of (a) EPCI values vs. experimental permeability values for carbonate samples. (b) EPCI values vs. simulated permeability estimates obtained using LBM and PNM simulations for sandstone samples.

Table 1 and Table 2 list EPCI values, permeability measurements, and permeability predictions obtained using the linear relationship between EPCI and measured permeability values. Carbonate samples are presented in Table 1, while the Sandstones are shown in Table 2. In the process of estimation, negative permeability estimates are set to zero. Average relative difference in predicting permeability values is 0.17 for carbonate samples, and 0.61 & 0.39 for LBM permeability and network permeability of sandstone samples, respectively. The increase in average error for the sandstone samples is likely to be caused by lower resolution of Sandstone samples. The resolution of carbonate samples is 1.0 μm/voxel, while the average resolution of Sandstone samples is about 6.0 µm/voxel. The EPCI index computes a rough estimate of the connectivity of pores. A slight error in measurements can have a larger impact on measurements error when compared with LBM simulations for example. These results show that in general the developed index generates reliable predictions for permeability values.

Table 1 Permeability Predictions for the Five Carbonate Samples. EPCI Values were Calibrated Using Experimental Data

| Sample | EPCI value | Measured permeability [mD] | EPCI Predictions | Relative difference between EPCI predictions and measurements | Average relative difference |
|--------|------------|----------------------------|------------------|---------------------------------------------------------------|-----------------------------|
| C1 | 0.01 | 1.80 | 1.86 | 0.04 | |
| C2 | 0.04 | 4.90 | 3.83 | 0.22 | |
| C3 | 0.11 | 7.00 | 10.13 | 0.45 | 0.17 |
| C4 | 0.16 | 16.00 | 13.62 | 0.15 | |
| C5 | 0.21 | 18.00 | 18.25 | 0.01 | |

Table 2 Permeability Predictions for the Nine Sandstone Samples. EPCI Values were Calibrated Using Permeability Estimates Obtained from LBM and PNM Simulations.

| Sample | EPCI | EPCI Predictions [mD] | | Relative difference between EPCI predictions and simulations | | Average relative difference | |
|---|---|---|---|---|---|---|---|
| | | LBM simulations | PNM simulations | LBM simulations | PNM simulations | LBM simulations | PNM simulations |
| S1 | 0.06 | 1216.91 | 1703.51 | 0.07 | 0.15 | 0.61 | 0.39 |
| S2 | 0.18 | 5562.13 | 6395.72 | 0.64 | 0.62 | | |
| S3 | 0.02 | 0.00 | 193.77 | 1.00 | 0.31 | | |
| S4 | 0.00 | 0.00 | 0.00 | 1.00 | 1.00 | | |
| S5 | 0.12 | 3253.98 | 3902.17 | 0.27 | 0.27 | | |
| S6 | 0.27 | 9010.98 | 10119.98 | 0.18 | 0.10 | | |
| S7 | 0.18 | 5727.23 | 6574.00 | 0.05 | 0.17 | | |
| S8 | 0.35 | 11726.73 | 13052.59 | 0.09 | 0.06 | | |
| S9 | 0.19 | 5916.38 | 6778.26 | 2.21 | 0.86 | | |

The developed algorithm achieves large computational savings when compared to LBM. For example, computing permeability using the EPCI values for an image volume of 300×300×300 is performed in 17.50 seconds. Estimating permeability using LBM takes around 1.67 hours. These computations were performed using a desktop computer equipped with two Intel Xeon (R) E5-2643 v2 processors operating at 3.50GHz and 128.00GB of installed RAM.

Furthermore, computing LBM values for large datasets require massive memory requirements. LBM simulations for a 3D data sample of size N×N×N require keeping track of about $N^3 \times d$ variables. Where d is the number of velocity directions and is normally chosen to be bigger than or equal to 15. In comparison, computing the proposed EPCI index require keeping track of $2N^2$ variables. Simulating LBM for samples of size 800×800×800 was not possible with the available desktop computer. In contrast, using the newly developed technique EPCI values were estimated in about 150 seconds.

## DISCUSSION AND FUTURE WORK

Experiments presented in this paper demonstrate that the EPCI index achieves good estimates for permeability values. The linear calibration used in this work require two parameters. It's worth noting that setting intercept to zero, slightly decreases the performance of the algorithm. For example, setting the intercept to zero in the Sandstone experiment, decreases the goodness of fit coefficient $R^2$ by about 0.01 for both LBM and PNM measurements. Relative error increases by 0.57 with LBM measurements and by 0.22 for PNM measurements.

The proposed index does not make explicit use of image resolution. We believe that this occurs for two reasons. First, image dimensions are expected to be equal for all rock volumes used in calibration. Second, in permeability studies, image resolution is set to value that best resolves the pore network structure. We believe that this relationship and use of constant image dimensions are the reasons behind EPCI's performance despite not taking resolution into account.

EPCI does not track the variations in cross-sectional areas in pore channels within the rock sample. When rock samples are imaged at proper resolutions, they show a variety of wide and narrow pore channels. EPCI assumes constant flow. This assumption underestimates flow in wide channels and overestimates flow in narrow channels. The effect of not calculating cross-sectional areas is minimized as the over- and under- estimation operations negate each other.

In the future, we plan to further validate the performance of the EPCI index by using comparisons with other permeability estimation techniques (Saxena et al., 2017). Sensitivity to segmentation artifacts and image resolution can also be incorporated in such study (Saxena et al., 2018). Development of an efficient method for tracking cross-sectional areas in the algorithm is expected to increase the performance of the developed index. The use of topological measures and Minkowski functionals (Liu et al., 2017, McClure et al., 2018) can be a possible way of integrating cross-sectional areas and other missing topological factors.

## CONCLUSIONS

In this paper, we introduce a new and fast algorithm to predict absolute permeability from X-ray CT images. The developed technique relies on the use of simple logical operations to track the connectivity of pores within an image volume. Our experiments demonstrate that predictions made by the new tool are in agreement with predictions made using laboratory measurements and LBM/PNM methods. The developed permeability estimation tool is computationally efficient and offers significant savings in computational time and memory requirements.